\documentclass[conference]{IEEEtran}
\IEEEoverridecommandlockouts

\usepackage{color}
\usepackage{graphicx}
\usepackage{amsmath}
\usepackage{amssymb}

\newcommand{\be}{\begin{equation}}
\newcommand{\ee}{\end{equation}}
\newcommand{\ba}{\begin{array}}
\newcommand{\ea}{\end{array}}

\renewcommand{\thesubsubsection}{\thesubsection.\arabic{subsubsection}}

\title{Hybrid Precoder and Combiner Design with One-Bit Quantized Phase Shifters in mmWave MIMO Systems}

\author{ \IEEEauthorblockN{Zihuan Wang$^{\dag}$, Ming Li$^{\dag}$,  Xiaowen Tian$^{\dag}$,  and Qian Liu$^{ \ddag}$
\vspace{-0.0 cm} }\\
\IEEEauthorblockA{$^{\dag}$School of Information and Communication Engineering   \\  Dalian University of Technology, Dalian, Liaoning 116024, China  \\
E-mail: \texttt{\{wangzihuan, tianxw\}@mail.dlut.edu.cn, mli@dlut.edu.cn}}

\IEEEauthorblockA{$^{\ddag}$  School of Computer Science and Technology \\  Dalian University of Technology, Dalian, Liaoning 116024, China \\ E-mail: \texttt{qianliu@dlut.edu.cn}} }

\begin{document}

\pagestyle{empty}

 \maketitle

\begin{abstract}
Analog/digital hybrid precoder and combiner have been widely used in millimeter wave (mmWave) multiple-input multiple-output (MIMO) systems due to its energy-efficient and economic superiorities. Infinite resolution of phase shifters (PSs) for the analog beamformer can achieve very close performance compared to the full-digital scheme but will result in high complexity and intensive power consumption. Thus, more cost effective and energy efficient low resolution PSs are typically used in practical mmWave MIMO systems. In this paper, we consider the joint hybrid precoder and combiner design with one-bit quantized PSs in mmWave MIMO systems. We propose to firstly design the analog precoder and combiner pair for each data stream successively, aiming at conditionally maximizing the spectral efficiency. We present a novel binary analog precoder and combiner optimization algorithm under a Rank-1 approximation of the interference-included equivalent channel with lower than quadratic complexity. Then the digital precoder and combiner are computed based on the obtained baseband effective channel to further enhance the spectral efficiency.  Simulation results demonstrate that the proposed algorithm outperforms the existing one-bit PSs based hybrid beamforming scheme.
\end{abstract}


\maketitle

\section{Introduction}

Millimeter wave (mmWave) communications, operating in the frequency bands from 30-300 GHz, have initiated a new era of wireless communication since they can solve the spectrum congestion problem thanks to the significantly large and unexploited mmWave frequency bands \cite{Pi CM 11}-\cite{Rappaport IA 13}. On the other hand, the mmWave communications with high frequencies enable a large antenna array in massive multiple-input
multiple-output (MIMO) systems to be packed in a small physical dimension \cite{Heath 16}. The large antenna array can provide sufficient gain by precoding and combining to overcome the severe free-space pathloss of mmWave channel. For MIMO systems operating in conventional cellular frequency bands, full-digital precoder and combiner are realized using a large number of expensive radio frequency (RF) chains and energy-intensive analog-to-digital converters (ADCs), which are impractical in the mmWave communication systems due to much higher carrier frequency and wider bandwidth.
Recently, analog/digital hybrid precoding and combining structures have emerged as a promising solution.

The hybrid precoding approach applies a large number of phase shifters (PSs) to implement high-dimensional analog precoder and a small number of RF chains for low-dimensional digital precoder to provide the necessary flexibility to perform multiplexing/multiuser techniques.
The existing hybrid beamforming schemes typically assume the infinite resolution of PSs with constant modulus constraints of the analog precoder \cite{Gao JSAC 16}-\cite{Sohrabi 16}, which can achieve extremely close performance compared to the full-digital case. However, it is of high complexity to realize accurate phases for PSs. According to the special characteristic of a mmWave channel, codebook-based analog beamformer design algorithms are also widely used to reduce complexity \cite{Ayach TWC 14}-\cite{Ayach Glob 13}, in which the columns of the analog precoder are selected from certain candidate vectors, such as array response vectors of the channel and discrete fourier transform (DFT) beamformers. Nevertheless, due to the high resolution of beam angles, it is still difficult to be implemented on the hardware. In an effort to further reduce the hardware costs and complexity, low resolution quantized PSs have been considered in the hybrid architectures. In \cite{Sohrabi 16}, the authors propose a hybrid beamforming algorithm using low resolution PSs. However, the performance is not very satisfactory with one-bit quantized PSs.

\begin{figure*}[!t]
\centering
\includegraphics[width= 6.2 in]{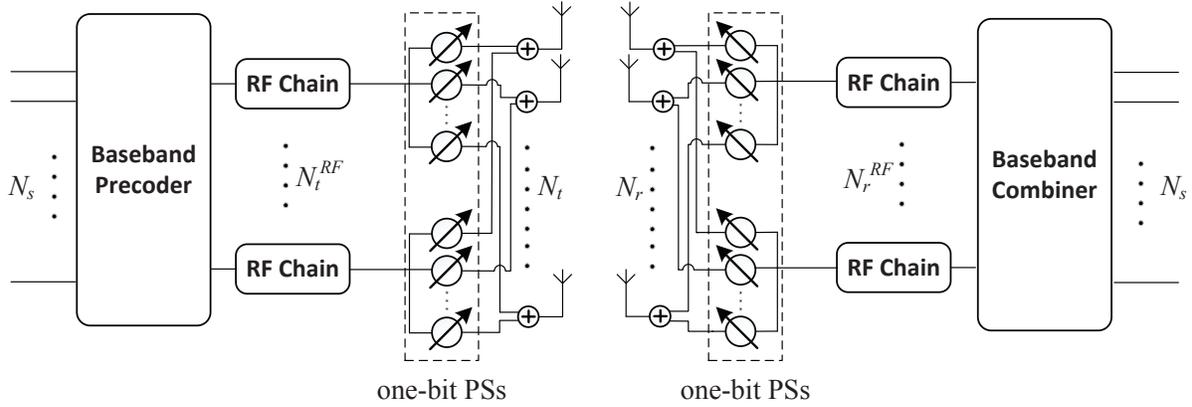}
\caption{The mmWave MIMO system using hybrid precoder and combiner with one-bit quantized PSs.}\label{fig:system_model}\vspace{-0.4 cm}
\end{figure*}

In this paper, we consider the hybrid precoder and combiner design with one-bit quantized PSs in mmWave MIMO systems. We propose to successively design the analog precoder and combiner pair for each data stream one by one, aiming at conditionally maximizing the spectral efficiency. Inspired by \cite{Karystinos 07}, we present a novel binary analog precoder and combiner optimization algorithm under a Rank-1 approximation of the interference-included equivalent channel with lower than quadratic complexity. Then, the digital precoder and combiner are computed based on the obtained baseband effective channel to further maximize the spectral efficiency. Simulation results demonstrate that the proposed algorithm can offer satisfactory performance improvement compared to the existing hybrid beamforming scheme with one-bit PSs.

The following notation is used throughout this paper. Boldface
lower-case letters indicate column vectors and boldface upper-case
letters indicate matrices; $\mathbb{C}$ denotes the set of all
complex numbers; $(\cdot)^T$ and $(\cdot)^H$  denote  the transpose and
transpose-conjugate operation, respectively;
$\mathbf{a}(i)$ denotes the $i$-th element of the vector $\mathbf{a}$.
$\mathbb{E} \{ \cdot \}$ represents
statistical expectation, $Re \{ \cdot \}$ extracts the real part of a complex number and $sign( \cdot )$ denotes the sign operator. Finally, $| \cdot |$, $\| \cdot \|$, and $\| \cdot \|_F$ are the scalar magnitude, vector norm, and Frobenius norm, respectively.

\section{System Model and Problem Formulation}
\label{sc:system model}

\subsection{System Model}

We consider a mmWave MIMO system using hybrid precoder and combiner with one-bit quantized PSs, as illustrated in Fig. \ref{fig:system_model}. The transmitter employs $N_t$ antennas and $N^{RF}_t$ RF chains to simultaneously transmit $N_s$ data streams to the receiver which is equipped with $N_r$ antennas and $N^{RF}_r$ RF chains. To ensure the efficiency of the communication with the limited number of RF chains, the number of data streams is constrained as $N_s = N_t^{RF} = N_r^{RF}$.

The transmitted symbols are firstly processed by a baseband digital precoder $\mathbf{F}_{BB} \in \mathbb{C}^{N_t^{RF}\times N_s}$, then up-converted to the RF domain via $N_t^{RF}$ RF chains before being precoded with an analog precoder $\mathbf{F}_{RF}$ of dimension $ N_t \times N_t^{RF}$. While the baseband precoder $\mathbf{F}_{BB}$ enables both amplitude and phase modifications, the analog precoder $\mathbf{F}_{RF}$ is assumed to have a constant amplitude $\frac{1}{\sqrt{N_t}}$ and one-bit quantized phases (i.e. binary phases) for each element, i.e. $\mathbf{F}_{RF} \in \frac{1}{\sqrt{N_t}}\{\pm1\}^{N_t\times N_t^{RF}}$.

The discrete-time transmitted signal can be written as
\begin{equation}
\mathbf{x} = \sqrt{P}\mathbf{F}_{RF} \mathbf{F}_{BB} \mathbf{s}
\end{equation}
where $\mathbf{s}$ is the $N_s \times 1$ symbol vector such that $\mathbb{E}\{ \mathbf{s} \mathbf{s}^H \} = \frac{1}{N_s}\mathbf{I}_{N_s}$, $P$ represents transmit power, and
 the total transmit power constraint is enforced
by normalizing $\mathbf{F}_{BB}$ such that $\|\mathbf{F}_{RF}\mathbf{F}_{BB}\|^2_F = N_s$.

For simplicity, we consider a narrowband block-fading propagation channel, which yields receive signal as
\begin{equation}
\mathbf{y} = \sqrt{P} \mathbf{H} \mathbf{F}_{RF}\mathbf{F}_{BB}\mathbf{s} + \mathbf{n} \label{eq:received signal 1}
\end{equation}
where $\mathbf{y}$ is the $N_r \times 1$ received vector, $\mathbf{H}$ is the $N_r \times N_t$ channel matrix, and  $\mathbf{n}\thicksim\mathcal{CN}(\mathbf{0},\sigma^2\mathbf{I}_{N_r})$ is the complex Gaussian noise vector corrupting the received signal.

The receiver uses its $ N_r^{RF}$ RF chains and
PSs to process the receive signal, and the processed signal has a form of
\begin{equation}
\mathbf{\widehat{s}} = \sqrt{P}\mathbf{W}^H_{BB}\mathbf{W}^H_{RF} \mathbf{H} \mathbf{F}_{RF}\mathbf{F}_{BB}\mathbf{s} + \mathbf{W}^H_{BB}\mathbf{W}^H_{RF}\mathbf{n} \label{eq:received signal 2}
\end{equation}
where $\mathbf{W}_{RF}$ is the $N_r \times N_r^{RF}$ analog combiner with similar constraints as $\mathbf{F}_{RF}$, $\mathbf{W}_{BB}$ is the $N_r^{RF} \times N_s$ digital baseband combiner and normalized to satisfy $\|\mathbf{W}_{RF}\mathbf{W}_{BB}\|^2_F = N_s$.


\subsection{Millimeter-Wave MIMO Channel Model}
The mmWave propagation in a massive MIMO system is well characterized by a limited spatial selectivity or scattering model, e.g. the Saleh-Valenzuela model, which allows us to accurately capture the mathematical structure in mmWave channels \cite{Ayach TWC 14}. The channel matrix $\mathbf{H}$ is assumed to be a sum contribution of $N_{cl}$ scattering clusters, each of which provides $N_{ray}$ propagation paths to the channel matrix $\mathbf{H}$. Therefore, the discrete-time narrow-band mmWave channel $\mathbf{H}$ can be formulated as
\begin{equation}
\mathbf{H}=\sqrt{\frac{N_t N_r}{N_{c1}N_{ray}}} \sum_{i=1}^{N_{cl}} \sum_{l=1}^{N_{ray}} \alpha_{il}\mathbf{a}_{r}(\theta_{il}^{r})\mathbf{a}_{t}(\theta_{il}^{t})^H
\end{equation}
where $\alpha_{il}\thicksim \mathcal{CN}(0,\sigma_{\alpha,i}^2)$ is the complex gain of the $l$-th propagation path (ray) in the $i$-th scattering cluster, following independent identically distributed (i.i.d.) form. Let $\sigma_{\alpha,i}^2$ represent the average power of the $i$-th cluster, and the total power satisfies $\sum_{i=1}^{N_{cl}}\sigma_{\alpha,i}^2 =N_{c1}$. $\theta_{il}^{t}$ and $\theta_{il}^{r}$ are the angle of departure (AoD) and the angle of arrival (AoA), respectively, which are assumed to be Laplacian-distributed with a mean cluster angle $\theta_{i}^{t}$ and $\theta_{i}^{r}$ as well as an angle spread of $\sigma_{\theta_i^t}$ and $\sigma_{\theta_i^r}$.
Finally, the array response vectors $\mathbf{a}_{r}(\theta^r)$ and $\mathbf{a}_{t}(\theta^t)$ are the antenna array
response vectors, which only depend on the antenna array structures. When the commonly used uniform linear arrays (ULAs) are considered, the receive antenna array response vector can be written as
\begin{equation}
\hspace{-0.0 cm}\mathbf{a}_r(\theta^r)=\frac{1}{\sqrt{N_r}}[1, {e}^{j\frac{2\pi}{\lambda}d\sin(\theta^r)}, \ldots , {e}^{j(N_r-1)\frac{2\pi}{\lambda}d\sin(\theta^r)}]^T  \hspace{-0.2 cm}
\end{equation}
where $\lambda$ is the signal wavelength, and $d$ is the distance between antenna elements. The transmit array response vector  $\mathbf{a}_t(\theta^t)$ can be written in a similar fashion.

\subsection{Problem Formulation}

We consider the problem of joint hybrid precoder and combiner design with one-bit quantized PSs in a mmWave MIMO system.
When Gaussian symbols are transmitted over the mmWave MIMO channel,
the achievable spectral efficiency is given by
\begin{eqnarray}
R \hspace{-0.2 cm} & = & \hspace{-0.2 cm} \mathrm{log}_2 \Bigg( \bigg|  \mathbf{I}_{N_s} + \frac{P}{N_s} \mathbf{R}_n^{-1}  \mathbf{W}^H_{BB}\mathbf{W}^H_{RF} \mathbf{H} \mathbf{F}_{RF}\mathbf{F}_{BB} \times   \nonumber \\ & & \hspace{2.2 cm}  \mathbf{F}_{BB}^H \mathbf{F}_{RF}^H \mathbf{H}^H \mathbf{W}_{RF}  \mathbf{W}_{BB}  \bigg| \Bigg), \label{eq:spectral efficiency}
\end{eqnarray}
where $\mathbf{R}_n \triangleq \sigma_n^2 \mathbf{W}^H_{BB}\mathbf{W}^H_{RF}\mathbf{W}_{RF}\mathbf{W}_{BB}$ is the noise covariance matrix after combining.

In this paper, we aim to design the digital beamformers $\mathbf{F}_{BB}$, $\mathbf{W}_{BB}$ as well as analog beamformers $\mathbf{F}_{RF}$, $\mathbf{W}_{RF}$ with the constant amplitude and one-bit quantized PSs to maximize the spectral efficiency:
\begin{equation}
\begin{aligned}
\Big\{\mathbf{F}_{RF}^\star, \mathbf{F}_{BB}^\star, \mathbf{W}_{RF}^\star, \mathbf{W}_{BB}^\star\Big\}= \textrm{arg}~\textrm{max} ~R\\
&\hspace{-4.6 cm}\textrm{s. t.}~~~~\mathbf{F}_{RF} \in \frac{1}{\sqrt{N_t}}\{\pm1\}^{N_t\times N_t^{RF}}, \\
&\hspace{-3.6 cm}\mathbf{W}_{RF} \in \frac{1}{\sqrt{N_r}}\{\pm1\}^{N_r\times N_r^{RF}},\\
&\hspace{-3.6 cm}\|\mathbf{F}_{RF}\mathbf{F}_{BB}\|^2_F = N_s,\\
&\hspace{-3.6 cm}\|\mathbf{W}_{RF}\mathbf{W}_{BB}\|^2_F = N_s.
\end{aligned}\label{eq:optimization problem}
\end{equation}
Obviously, the optimization problem (\ref{eq:optimization problem}) is a non-convex NP-hard problem. In the next section, we attempt to decompose the original problem into a series sub-problems and seek a sub-optimal solution with a satisfactory performance. 

\section{Proposed Hybrid Precoder and Combiner Design}
\label{sec:Proposed Design Full Connect}

In this section, we first focus on the analog precoder and combiner design. Then, having the baseband effective channel associated with the obtained optimal analog precoder and combiner, the digital precoder and combiner are computed to further maximize the spectral efficiency.

\subsection{Analog Precoder and Combiner Design}

Under high signal-to-noise-ratio (SNR) circumstance, the achievable spectral efficiency in (\ref{eq:spectral efficiency}) can be rewritten as
\begin{eqnarray}
R \hspace{-0.2 cm} & \approx & \hspace{-0.2 cm} \mathrm{log}_2 \Bigg( \bigg|  \frac{P}{N_s} \mathbf{R}_n^{-1}  \mathbf{W}^H_{BB}\mathbf{W}^H_{RF} \mathbf{H} \mathbf{F}_{RF}\mathbf{F}_{BB} \times   \nonumber \\ & & \hspace{1.6 cm}  \mathbf{F}_{BB}^H \mathbf{F}_{RF}^H \mathbf{H}^H \mathbf{W}_{RF}  \mathbf{W}_{BB}  \bigg| \Bigg).\label{eq:high SNR}
\end{eqnarray}
In addition, it has been verified \cite{Sohrabi 16} that for large-scale MIMO systems, the optimal analog beamformers are approximately orthogonal, i.e. $\mathbf{F}_{RF}^H\mathbf{F}_{RF} \propto \mathbf{I}$. This enables us to assume $\mathbf{F}_{BB}\mathbf{F}_{BB}^H\approx\zeta^2\mathbf{I}$ when $N_{RF}^t = N_s$, where $\zeta^2$ is a normalization factor. Similarly, we have $\mathbf{W}_{BB}\mathbf{W}_{BB}^H\approx\xi^2\mathbf{I}$ and $\mathbf{W}_{BB}^H\mathbf{W}_{RF}^H\mathbf{W}_{RF}\mathbf{W}_{BB}\approx\mathbf{I}$. Let $\gamma^2\triangleq\zeta^2\xi^2$, then (\ref{eq:high SNR}) can be further simplified as
\begin{align}
&\hspace{-0.3 cm}R\approx \mathrm{log}_2 \Bigg( \bigg|  \frac{P\gamma^2}{N_s\sigma^2}\mathbf{W}^H_{RF} \mathbf{H} \mathbf{F}_{RF} \mathbf{F}_{RF}^H \mathbf{H}^H \mathbf{W}_{RF} \bigg| \Bigg)\\
&\hspace{-0.1 cm}\overset{(a)}{=}\mathrm{log}_2 \left(\frac{P\gamma^2}{N_s\sigma^2}\right)+2\times \mathrm{log}_2 \Bigg( \bigg|\mathbf{W}^H_{RF} \mathbf{H} \mathbf{F}_{RF}\bigg| \Bigg)
\end{align}
where $(a)$ follows the fact that $|\mathbf{X}\mathbf{Y}|=|\mathbf{X}||\mathbf{Y}|$ when $\mathbf{X}$ and $\mathbf{Y}$ are both square matrices.

Therefore, the binary analog precoder and combiner design can be formulated as:
\begin{equation}
\begin{aligned}
\hspace{-0.1 cm}\left\{\mathbf{F}_{RF}^\star, \mathbf{W}_{RF}^\star \right\}= \textrm{arg}\hspace{-0.4 cm}\underset{\substack{\mathbf{F}_{RF} \in \frac{1}{\sqrt{N_t}}\{\pm 1\}^{N_t\times N_t^{RF}} \\ \mathbf{W}_{RF} \in \frac{1}{\sqrt{N_r}}\{\pm 1\}^{N_r\times N_r^{RF}}}}{\textrm{max}}\hspace{-0.2 cm}\mathrm{log}_2 \bigg( \Big|\mathbf{W}^H_{RF} \mathbf{H} \mathbf{F}_{RF}\Big| \bigg).
\end{aligned}\label{eq:objective function}
\end{equation}
Unfortunately, this binary analog beamformer design problem is still a NP-hard problem and has complexity of $\mathcal{O}(2^{N_tN_rN_t^{RF} N_r^{RF}})$. Hence, we propose to further decompose this difficult optimization problem into a series of sub-problems, in which each transmit/receive RF chain pair is considered one by one, and the analog precoder and combiner for each pair are successively designed.

In particular, we perform singular value decomposition (SVD)
\begin{equation}
\mathbf{H}=\mathbf{U}\mathbf{\Sigma}\mathbf{V}^H
\end{equation}
where $\mathbf{U}$ is an $N_t \times N_t$ unitary matrix, $\mathbf{V}$ is an $N_r \times N_r$ unitary matrix and $\mathbf{\Sigma}$ is a rectangular diagonal matrix of singular values arranged in a decreasing order on the diagonal. Utilizing the sparse nature of mmWave channel, we only retain the $N_s$ strongest components since they contain the most power of the channel. Let
$\mathbf{\widehat{U}}\triangleq\mathbf{U}(:,1:N_s)$, $\mathbf{\widehat{\Sigma}}\triangleq\mathbf{\Sigma}(1:N_s,1:N_s)$ and $\mathbf{\widehat{V}}\triangleq\mathbf{V}(:,1:N_s)$.
Then the objective in (\ref{eq:objective function}) can be converted into
\begin{eqnarray}
&&\hspace{-1.0 cm}\mathrm{log}_2 \bigg( \Big|\mathbf{W}^H_{RF} \mathbf{H} \mathbf{F}_{RF}\Big| \bigg)
\approx\mathrm{log}_2 \bigg( \Big|\mathbf{W}^H_{RF} \mathbf{\widehat{U}}\mathbf{\widehat{\Sigma}}\mathbf{\widehat{V}}^H \mathbf{F}_{RF}\Big| \bigg).\label{eq:decomposition}
\end{eqnarray}
Partition the analog precoding and combining matrices as $\mathbf{F}_{RF}=[\mathbf{F}_{RF,{N_s-1}}~\mathbf{f}_{RF,{N_s}}]$ and $\mathbf{W}_{RF}=[\mathbf{W}_{RF,{N_s-1}}~\mathbf{w}_{RF,{N_s}}]$, respectively. Then, the formulation (\ref{eq:decomposition}) can be further transformed in (\ref{eq:decomposition1})-(\ref{eq:transformation}), which are presented at the top of following page,
\begin{figure*}
\begin{small}
\begin{eqnarray}
&&\hspace{-1.0 cm}\mathrm{log}_2 \bigg( \Big|\mathbf{W}^H_{RF} \mathbf{\widehat{U}}\mathbf{\widehat{\Sigma}}\mathbf{\widehat{V}}^H \mathbf{F}_{RF}\Big| \bigg)=\mathrm{log}_2 \bigg( \Big|\mathbf{\widehat{\Sigma}}\mathbf{\widehat{V}}^H \mathbf{F}_{RF}\mathbf{W}^H_{RF} \mathbf{\widehat{U}}\Big| \bigg)=\mathrm{log}_2 \Bigg( \bigg|\mathbf{\widehat{\Sigma}}\mathbf{\widehat{V}} ^H \left[\mathbf{F}_{RF,{N_s-1}}~\mathbf{f}_{RF,{N_s}}\right]\left[\mathbf{W}_{RF,{N_s-1}} \mathbf{w}_{RF,{N_s}}\right]^H \mathbf{\widehat{U}}\bigg| \Bigg)\label{eq:decomposition1}\\
&&\hspace{-0.8 cm}=\mathrm{log}_2 \Bigg( \bigg|\mathbf{\widehat{\Sigma}}\mathbf{\widehat{V}}^H \mathbf{F}_{RF,{N_s-1}}\mathbf{W}_{RF,{N_s-1}}^H\mathbf{\widehat{U}}+ \mathbf{\widehat{\Sigma}}\mathbf{\widehat{V}}^H\mathbf{f}_{RF,{N_s}}\mathbf{w}_{RF,{N_s}}^H \mathbf{\widehat{U}} \bigg| \Bigg)\\
&&\hspace{-0.8 cm}\approx\mathrm{log}_2 \Bigg( \bigg|\left(\mathbf{\widehat{\Sigma}}\mathbf{\widehat{V}}^H \mathbf{F}_{RF,{N_s-1}}\mathbf{W}_{RF,{N_s-1}}^H\mathbf{\widehat{U}}\right)\Big[ \mathbf{I}+\left(\alpha\mathbf{I}+\mathbf{\widehat{\Sigma}}\mathbf{\widehat{V}}^H \mathbf{F}_{RF,{N_s-1}}\mathbf{W}_{RF,{N_s-1}}^H\mathbf{\widehat{U}}\right)^{-1} \mathbf{\widehat{\Sigma}}\mathbf{\widehat{V}}^H\mathbf{f}_{RF,{N_s}}\mathbf{w}_{RF,{N_s}}^H \mathbf{\widehat{U}}\Big] \bigg| \Bigg)\\
&&\hspace{-0.8 cm}=\mathrm{log}_2 \Bigg( \bigg|\left(\mathbf{\widehat{\Sigma}}\mathbf{\widehat{V}}^H \mathbf{F}_{RF,{N_s-1}}\mathbf{W}_{RF,{N_s-1}}^H\mathbf{\widehat{U}}\right)\bigg| \Bigg)+
\mathrm{log}_2 \Bigg( \bigg|\Big[ \mathbf{I}+\left(\alpha\mathbf{I}+\mathbf{\widehat{\Sigma}}\mathbf{\widehat{V}}^H \mathbf{F}_{RF,{N_s-1}}\mathbf{W}_{RF,{N_s-1}}^H\mathbf{\widehat{U}}\right)^{-1} \mathbf{\widehat{\Sigma}}\mathbf{\widehat{V}}^H\mathbf{f}_{RF,{N_s}}\mathbf{w}_{RF,{N_s}}^H \mathbf{\widehat{U}}\Big] \bigg| \Bigg)\label{eq:transformation}
\end{eqnarray}
\end{small}
\hrule \vspace{-0.4 cm}
\end{figure*}
where $\alpha$ is a very small scalar to assure invertibility. Note that the first term $\mathrm{log}_2 \left( \left|\left(\mathbf{\widehat{\Sigma}}\mathbf{\widehat{V}}^H \mathbf{F}_{RF,{N_s-1}}\mathbf{W}_{RF,{N_s-1}}^H\mathbf{\widehat{U}}\right)\right| \right)$ in (\ref{eq:transformation}) can be further decomposed using a similar procedure. Thus, the objective in (\ref{eq:decomposition}) can be reformulated as a summation of the conditionally achievable spectral efficiency associated with each analog beamformer pair:
\begin{eqnarray}
&&\hspace{-1.2 cm}\mathrm{log}_2 \bigg( \Big|\mathbf{W}^H_{RF} \mathbf{H} \mathbf{F}_{RF}\Big| \bigg) \approx\sum_{l=1}^{N_s}\mathrm{log}_2 \Bigg( \bigg|\Big[ \mathbf{I}+\big(\alpha\mathbf{I}+\nonumber \\ && \hspace{-0.7 cm}\mathbf{\widehat{\Sigma}}\mathbf{\widehat{V}}^H \mathbf{F}_{RF,{l-1}} \mathbf{W}_{RF,{l-1}}^H\mathbf{\widehat{U}}\big)^{-1}\mathbf{\widehat{\Sigma}}\mathbf{\widehat{V}}^H\mathbf{f}_{RF,{l}}\mathbf{w}_{RF,{l}}^H \mathbf{\widehat{U}}\Big] \bigg| \Bigg) \\
&&\hspace{-0.8 cm}=\sum_{l=1}^{N_s}\mathrm{log}_2 \bigg(1+\mathbf{w}_{RF,{l}}^H \mathbf{\widehat{U}}\Big(\alpha\mathbf{I}+ \nonumber \\
&&\hspace{0.4 cm}\mathbf{\widehat{\Sigma}}\mathbf{\widehat{V}}^H \mathbf{F}_{RF,{l-1}}\mathbf{W}_{RF,{l-1}}^H\mathbf{\widehat{U}}\Big)^{-1} \mathbf{\widehat{\Sigma}}\mathbf{\widehat{V}}^H\mathbf{f}_{RF,{l}}\bigg)\label{eq:final}
\end{eqnarray}
where $\mathbf{F}_{RF,0}\triangleq\mathbf{0}$ and $\mathbf{W}_{RF,0}\triangleq\mathbf{0}$.
According to (\ref{eq:final}), we notice that the total achievable spectral efficiency problem can be decomposed into a series sub-problems, each of which only considers a corresponding pair of precoder $\mathbf{f}_{RF,l}$ and combiner $\mathbf{w}_{RF,l}$. Therefore, the analog precoder and combiner pair can be successively designed one by one.

Denote $\mathbf{Q}_1\triangleq\mathbf{H}$ and $\mathbf{Q}_l\triangleq \mathbf{w}_{RF,{l}}^H \mathbf{\widehat{U}}(\alpha\mathbf{I}+\mathbf{\widehat{\Sigma}}\mathbf{\widehat{V}}^H \mathbf{F}_{RF,{l-1}} \mathbf{W}_{RF,{l-1}}^H\mathbf{\widehat{U}})^{-1} \mathbf{\widehat{\Sigma}}\mathbf{\widehat{V}}^H\mathbf{f}_{RF,{l}},~l=2, \ldots , N_s$.
For the $l$-th analog beamformer pair, we have
\begin{equation}
\begin{aligned}
 \left\{\mathbf{f}_{{RF},l}^{\star},\mathbf{w}_{{RF},l}^{\star}  \right\}= \textrm{arg}\underset{\substack{\mathbf{f}_{{RF},l}  \in \frac{1}{\sqrt{N_t}}\{\pm 1\}^{N_t}\\ \mathbf{w}_{{RF},l}  \in \frac{1}{\sqrt{N_r}}\{\pm 1\}^{N_r}}}{\textrm{max}} \left| \mathbf{w}_{RF,l}^H \mathbf{Q}_l\mathbf{f}_{RF,l} \right|.
\label{eq:beam sel}
\end{aligned}
\end{equation}
It is noted that this optimization problem can be solved through exhaustive search with the complexity $\mathcal{O}(2^{N_t N_r})$. To further reduce the complexity, we perform SVD on $\mathbf{Q}_l$ as
\begin{equation}
\mathbf{Q}_l=\sum_{i=1}^{N_s-l+1}\lambda_{l,i}\mathbf{p}_{l,i}\mathbf{q}_{l,i}^H,~\lambda_{l,1}\geq\lambda_{l,2} \geq\ldots\geq\lambda_{l,{N_s-l+1}},
\end{equation}
where $\mathbf{p}_{l,i}$ and $\mathbf{q}_{l,i}$ are the $i$-th left and right singular vectors of matrix $\mathbf{Q}_l$, respectively, $\lambda_{l,i}$ is the $i$-th largest singular value. Then, the objective in (\ref{eq:beam sel}) can be rewritten as
$| \mathbf{w}_{RF,l}^H \mathbf{Q}_l\mathbf{f}_{RF,l} |=| \sum_{i=1}^{N_s-l+1}\lambda_{l,i}\mathbf{w}_{RF,l}^H \mathbf{p}_{l,i} \mathbf{q}_{l,i}^H\mathbf{f}_{RF,l} |$.
If we simplify the optimization problem by keeping only the strongest term, i.e. $\mathbf{Q}_l\approx\lambda_{l,1}\mathbf{p}_{l,1}\mathbf{q}_{l,1}^H$, the optimization function in (\ref{eq:beam sel}) can be described by
\begin{equation}
\begin{aligned}
\hspace{-0.1 cm}\left\{\mathbf{f}_{{RF},{l}}^{\star},\mathbf{w}_{{RF},{l}}^\star  \right\}= \textrm{arg}\hspace{-0.3 cm}\underset{\substack{\mathbf{f}_{{RF},l}  \in \frac{1}{\sqrt{N_t}}\{\pm 1\}^{N_t}\\ \mathbf{w}_{{RF},l}  \in \frac{1}{\sqrt{N_r}}\{\pm 1\}^{N_r}}}{\textrm{max}}\hspace{-0.2 cm} \left| \lambda_{l,1}\mathbf{w}_{RF,l}^H \mathbf{p}_{l,1} \mathbf{q}_{l,1}^H\mathbf{f}_{RF,l} \right|.
\end{aligned}\label{eq:joint optimize}
\end{equation}
Now, this joint optimization problem (\ref{eq:joint optimize}) is equivalent to individually design the analog precoder and combiner:
\begin{eqnarray}
&\mathbf{f}_{{RF},{l}}^{\star}=\underset{\mathbf{f}_{{RF},l}  \in \frac{1}{\sqrt{N_t}}\{\pm1\}^{N_t}}{\textrm{max}}  \left| \mathbf{f}_{RF,l}^H \mathbf{q}_{l,1} \right|,\label{eq:analog precoder design1}\\
&\mathbf{w}_{{RF},{l}}^{\star}=\underset{\mathbf{w}_{{RF},l}  \in \frac{1}{\sqrt{N_r}}\{\pm1\}^{N_r}}{\textrm{max}}  \left| \mathbf{w}_{RF,l}^H \mathbf{p}_{l,1} \right|.
\end{eqnarray}
We can observe that these two simplified optimization problems still have complexity exponential in the number of antennas, which are difficult to be solved directly.
Next, we attempt to construct a candidate beamformer set, from which the optimal beamformer can be selected with lower than quadratic complexity. The analog precoder design is considered as an example and the ananlog combiner design follows the same method.

We first introduce an auxiliary variable $\phi \in [-\pi,\pi)$ and rewrite maximization problem (\ref{eq:analog precoder design1}) as:
\begin{equation}
\begin{aligned}
&\hspace{-0.2 cm}\left\{\phi^\star, \mathbf{f}_{RF,l}^\star\right\}=\textrm{arg}\underset{\substack{\phi \in [-\pi,\pi)\\ \mathbf{f}_{{RF},l}  \in \frac{1}{\sqrt{N_t}}\{\pm1\}^{N_t}}}{\textrm{max}}   Re\left\{\mathbf{f}_{RF,l}^H \mathbf{q}_{l,1} e^{-j\phi}\right\} \\
&\hspace{0.2 cm}=\textrm{arg}\underset{\substack{\phi \in [-\pi,\pi)\\ \mathbf{f}_{{RF},l}  \in \frac{1}{\sqrt{N_t}}\{\pm1\}^{N_t}}}{\textrm{max}}   \sum_{i=1}^{N_t}\mathbf{f}_{RF,l}(i) |\mathbf{q}_{l,1}(i)| \cos(\phi-\varphi_i)
\end{aligned}\label{eq:analog precoder design2}
\end{equation}
where $\varphi_i$ indicates the phase of the $i$-th element of $\mathbf{p}_{l,1}$. Obviously, given any $\phi$, the conditionally optimal analog precoder can be obtained by $\mathbf{f}_{RF,l}(i)=\frac{1}{\sqrt{N_t}}sign\left(\cos\left(\phi-\varphi_i\right)\right), i=1, \ldots , N_t$, and the problem (\ref{eq:analog precoder design2}) can be transformed into finding the optimal $\phi^\star$ \vspace{-0.2 cm}
\begin{equation}
\begin{aligned}
&\phi^\star=\textrm{arg}\underset{\phi \in [-\pi,\pi)}{\textrm{max}}  \sum_{i=1}^{N_t} |\mathbf{q}_{l,1}(i)| |\cos(\phi-\varphi_i)|.
\end{aligned}\label{eq:candidate generate}
\end{equation}
To find the optimal $\phi^\star$, we first define the angles $\widehat{\varphi_i}$ as
\begin{equation}
\widehat{\varphi_i}\triangleq\left\{
\begin{aligned}
 & \varphi_i-\pi, ~~~\varphi_i \in \left[\frac{\pi}{2},\frac{3\pi}{2}\right],~~i=1, \ldots, N_t,\\
  &\varphi_i,~~~~~~~~\varphi_i \in \left[-\frac{\pi}{2},\frac{\pi}{2}\right),~~i=1, \ldots, N_t.
\end{aligned}
\right.
\end{equation}
Then, we map the angles $\widehat{\varphi}_i, ~i=1, \ldots , N_t$, to $\widetilde{\varphi}_i, ~i=1, \ldots , N_t$, which is rearranged in ascending order, i.e. $\widetilde{\varphi}_1 \leq \widetilde{\varphi}_2 \leq \ldots \leq \widetilde{\varphi}_{N_t} \in [-\frac{\pi}{2},\frac{\pi}{2})$.
Note that in (\ref{eq:candidate generate}), the maximization problem associated with $\phi$ can be solved over any interval of length $\pi$. Then, we can assume the optimal $\phi$ is in the interval of $[\widetilde{\varphi}_1-\frac{\pi}{2},\widetilde{\varphi}_1+\frac{\pi}{2})$.
Considering $N_t$ non-overlapping subintervals $\left[\widetilde{\varphi}_1-\frac{\pi}{2}, \widetilde{\varphi}_2-\frac{\pi}{2}\right),\left[\widetilde{\varphi}_2-\frac{\pi}{2}, \widetilde{\varphi}_3-\frac{\pi}{2}\right),\ldots,\left[\widetilde{\varphi}_{N_t}-\frac{\pi}{2}, \widetilde{\varphi}_1+\frac{\pi}{2}\right)$ in $[\widetilde{\varphi}_1-\frac{\pi}{2},\widetilde{\varphi}_1+\frac{\pi}{2})$, the previous problem (\ref{eq:candidate generate}) can be solved by examining each subinterval separately. For $\phi$ in the $i$-th subinterval, the corresponding optimal analog precoder has a form of
\begin{equation}
\mathbf{\widetilde{f}}_{l,i}=[\underbrace{1 \ldots 1}_i\underbrace{-1 \ldots -1}_{N_t-i}]^T \nonumber
\end{equation}
Since the optimal $\phi^\star$ must be included in one of these subintervals, we can construct a beamformer set $\mathcal{\widetilde{F}}_l=\{\mathbf{\widetilde{f}}_{l,1}, \ldots, \mathbf{\widetilde{f}}_{l,{N_t}}\}$ associated with all the subintervals without loss of optimality.
After that, given the inverse mapping from $\widetilde{\varphi}$ to $\widehat{\varphi}$, we rearrange the corresponding elements of
$\mathbf{\widetilde{f}}_{l,k},\forall k=1, \ldots , N_t$ to $\mathbf{\widehat{f}}_{l,k},\forall k=1, \ldots , N_t$, resulting in $\widehat{\mathcal{F}}_l=\{\mathbf{\widehat{f}}_{l,1}, \ldots, \mathbf{\widehat{f}}_{l,{N_t}}\}$. Finally, the candidate beamformer set $\mathcal{F}_l$ is defined by $\mathcal{F}_l=\{\mathbf{f}_{l,1}, \ldots, \mathbf{f}_{l,{N_t}}\}$ and
\begin{equation}
\mathbf{f}_{l,k}(i)\triangleq\left\{
\begin{aligned}
 &-\mathbf{\widehat{f}}_{l,k}(i), ~~~\varphi_i \in \left[\frac{\pi}{2},\frac{3\pi}{2}\right],~\forall k=1, \ldots , N_t,\\
  &~~\mathbf{\widehat{f}}_{l,k}(i),~~~~\varphi_i \in \left[-\frac{\pi}{2},\frac{\pi}{2}\right),~\forall k=1, \ldots , N_t.
\end{aligned}
\right.\label{eq:code-boook}
\end{equation}
Similarly, we can also construct a candidate set $\mathcal{W}_l$ for analog combiner design.
Finally, with the candidate beamformer sets $\mathcal{F}_l$ and $\mathcal{W}_l$, the analog precoder and combiner design issue (\ref{eq:beam sel}) becomes
\begin{eqnarray}
&\left\{\mathbf{f}_{{RF},{l}}^\star,\mathbf{w}_{{RF},{l}}^\star  \right\}=\textrm{arg}\underset{ \substack{\mathbf{f}_{{RF},l}  \in \mathcal{F}_l\\ \mathbf{w}_{{RF},l}  \in \mathcal{W}_l}}{\textrm{max}}   \left| \mathbf{w}_{RF,l}^H \mathbf{Q}_l\mathbf{f}_{RF,l} \right|,\nonumber \\ &\hspace{0.6 cm}l=1, \ldots , N_s.
\end{eqnarray}

\subsection{Digital Precoder and Combiner Design}

After all the analog beamformer pairs have been determined, we can obtain the baseband effective channel
\begin{equation}
\mathbf{\widetilde{H}} \triangleq \left(\mathbf{W}_{RF}^\star\right)^H \mathbf{H} \mathbf{F}_{RF}^\star,  \label{eq:effective channel}
\end{equation}
where $\mathbf{F}_{RF}^\star \triangleq [\mathbf{f}_{RF,1}^\star, \ldots, \mathbf{f}_{RF,{N_s}}^\star] $ and $\mathbf{W}_{RF}^\star\triangleq [\mathbf{w}_{RF,1}^\star, \ldots, \mathbf{w}_{RF,{N_s}}^\star]$. For baseband precoder and combinder design, we perform SVD
\begin{equation}
\mathbf{\widetilde{H}} = \mathbf{C} \mathbf{D} \mathbf{S}^H,
\end{equation}
where $ \mathbf{C}$ and $\mathbf{S}$ are $N_s \times N_s$ unitary matrices, $\mathbf{D}$ is an $N_s \times N_s$ diagonal matrix of singular values. Then, an SVD-based baseband digital precoder and combiner are employed to further enhance the spectral efficiency:
\begin{eqnarray}
 \mathbf{F}^\star_{{BB}}  =  \mathbf{S}, \\
\mathbf{W}^\star_{{BB}}  =  \mathbf{C}.
\end{eqnarray}
Finally, we normalize baseband precoder and combiner by
\begin{align}
&\mathbf{F}^\star_{{BB}}=\frac{\sqrt{N_s}\mathbf{F}^\star_{{BB}}} {\|\mathbf{F}^\star_{RF}\mathbf{F}^\star_{{BB}}\|_F},\\
&\mathbf{W}^\star_{{BB}}=\frac{\sqrt{N_s}\mathbf{W}^\star_{{BB}}} {\|\mathbf{W}^\star_{RF}\mathbf{W}^\star_{{BB}}\|_F}.
\end{align}

\vspace{0.1 cm}
\section{Simulation Results}
\label{sc:Simulation}

In this Section, we illustrate the simulation results of the proposed joint hybrid precoder and combiner design. Transmitter and receiver are equipped with a $64$-antenna ULA and a $16$-antenna ULA, respectively, where antenna spacing is $d=\frac{\lambda}{2}$. The numbers of RF chains at transmitter and receiver are $N_{t}^{RF} = N_{r}^{RF} = 4$, so is the number of data streams $N_s = 4$.  The channel parameters are set as $N_{cl} = 10$ clusters, $N_{ray} = 10$ rays per cluster, and the
average power of the $i$-th cluster is $ \sigma^2_{\alpha,i} = c\frac{7}{10}^i$ where $c=(\sum_{i=1}^{N_{cl}}(\frac{7}{10})^i)^{-1}N_{cl}$. The azimuths of the AoAs/AoDs within a cluster are assumed to be Laplacian-distributed with an angle spread of $\sigma_{\theta_i^r}=\sigma_{\theta_i^t}=2.5^\circ$. The mean cluster AoDs are assumed to be uniformly distributed over $[0,2\pi]$, while the mean cluster AoAs are uniformly distributed over an arbitrary $\frac{\pi}{3}$ sector.

\begin{figure}[!t]
\centering

  \includegraphics[width=3.1 in]{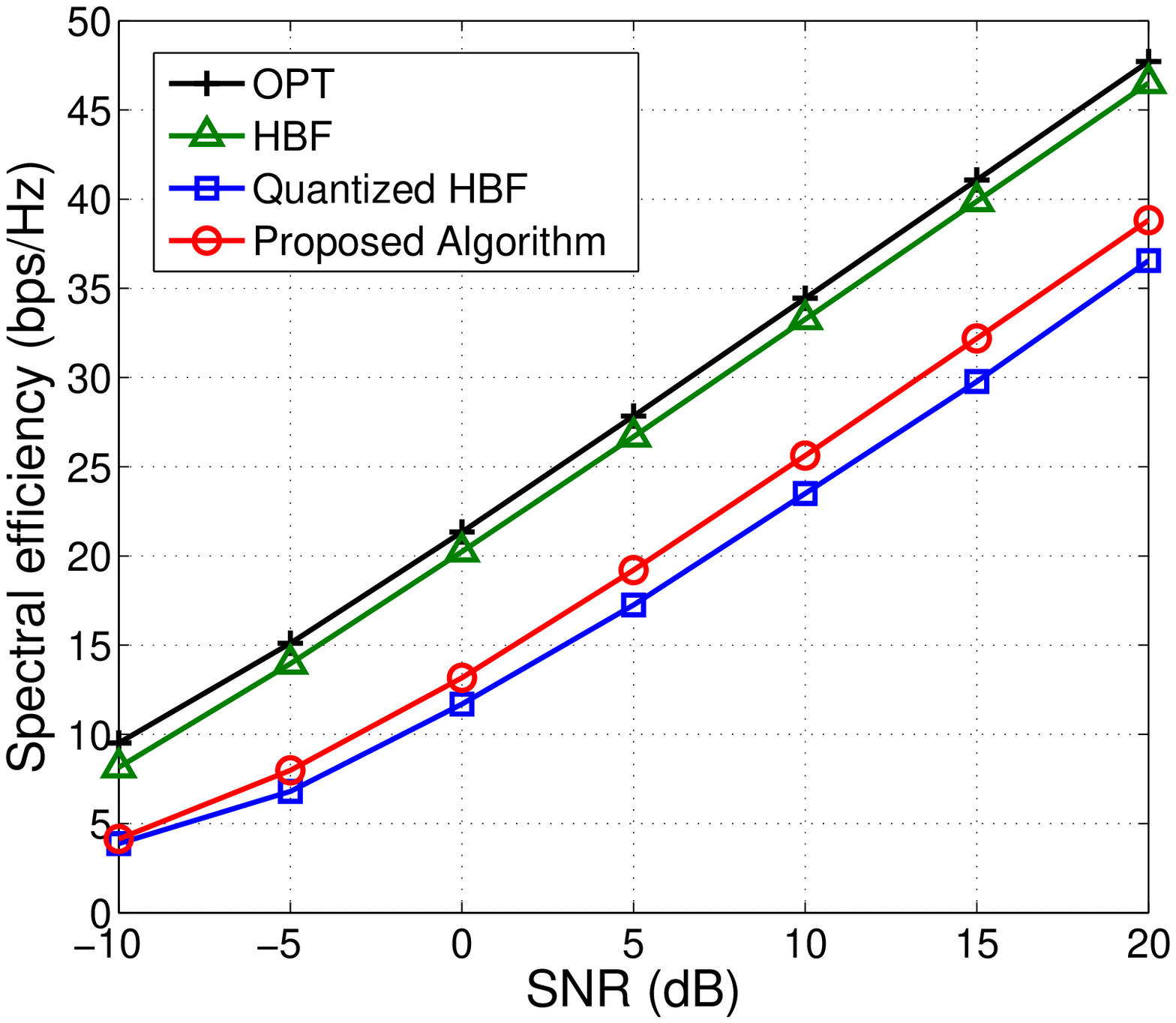}
  \vspace{-0.3 cm}
  \caption{Spectral efficiency versus SNR ($N_t=64$, $N_r=16$, $N_t^{RF}= N_r^{RF}=4$, $N_s=4$).}\label{fig:rate_vs_snr_M64N16Ns4}

  \includegraphics[width=3.1 in]{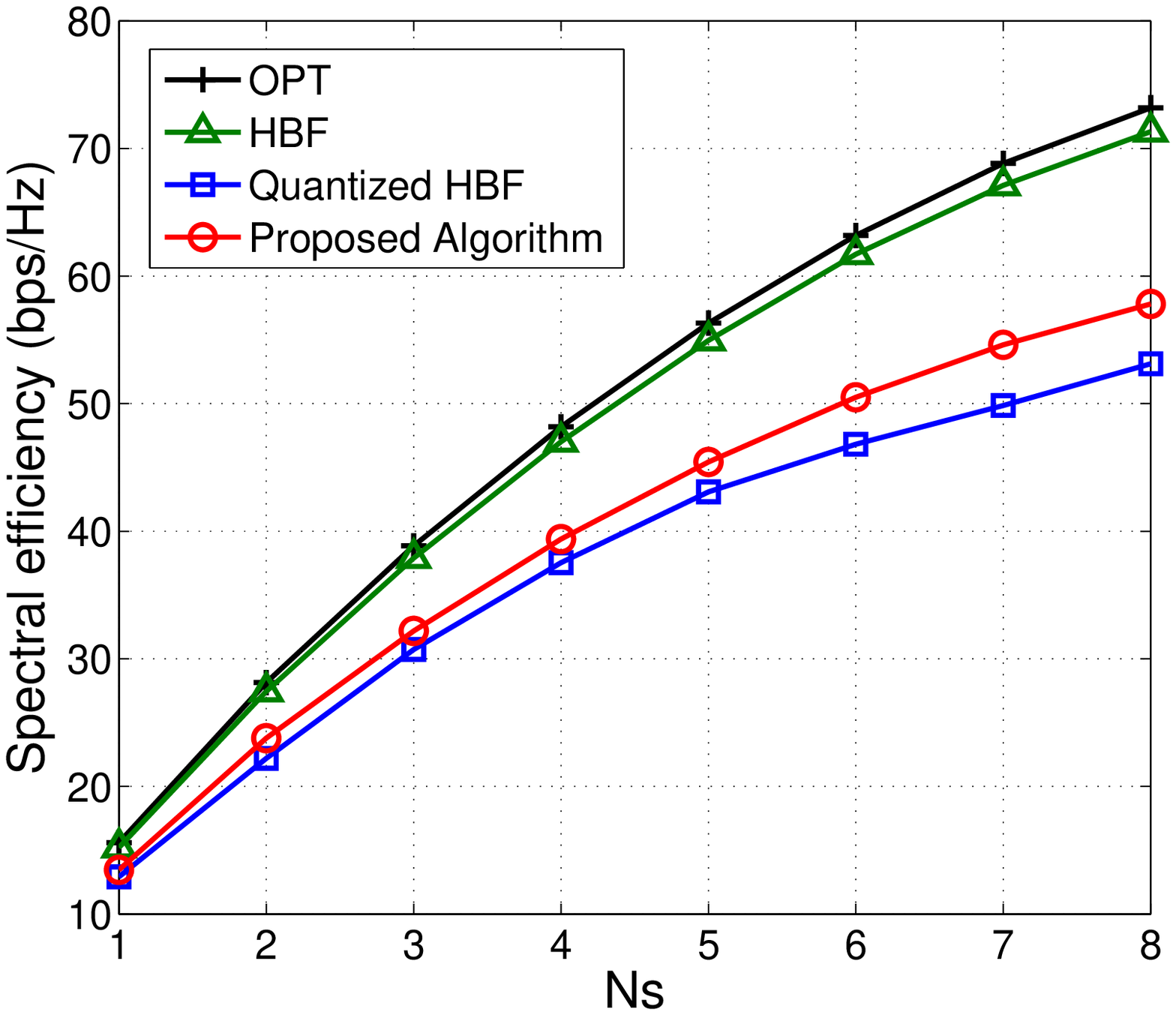}
  \vspace{-0.3 cm}
  \caption{Spectral efficiency versus $N_s$ ($N_t=64, N_r=16$, $N_t^{RF}= N_r^{RF}=N_s$, SNR = $20$dB).}\label{fig:rate_vs_Ns}\vspace{-0.4 cm}

\end{figure}

Fig. \ref{fig:rate_vs_snr_M64N16Ns4} shows the spectral efficiency versus SNR over $10^6$ channel realizations. For the comparison purpose, we also include the state-of-the-art algorithm introduced in \cite{Sohrabi 16}, in which a hybrid beamforming (HBF) with infinite resolution PSs is first proposed, then an one-bit quantized version (Quantized HBF) is presented. To the best of our knowledge, the Quantized HBF approach can achieve the best performance using one-bit quantized PSs. The optimal (OPT) full-digital beamforming scheme with the unconstrained SVD algorithm is also plotted as the performance benchmark. It can be observed that our proposed algorithm can achieve a satisfactory performance and outperform the Quantized HBF scheme. 

\begin{figure}[!t]
\centering
\includegraphics[width=3.1 in]{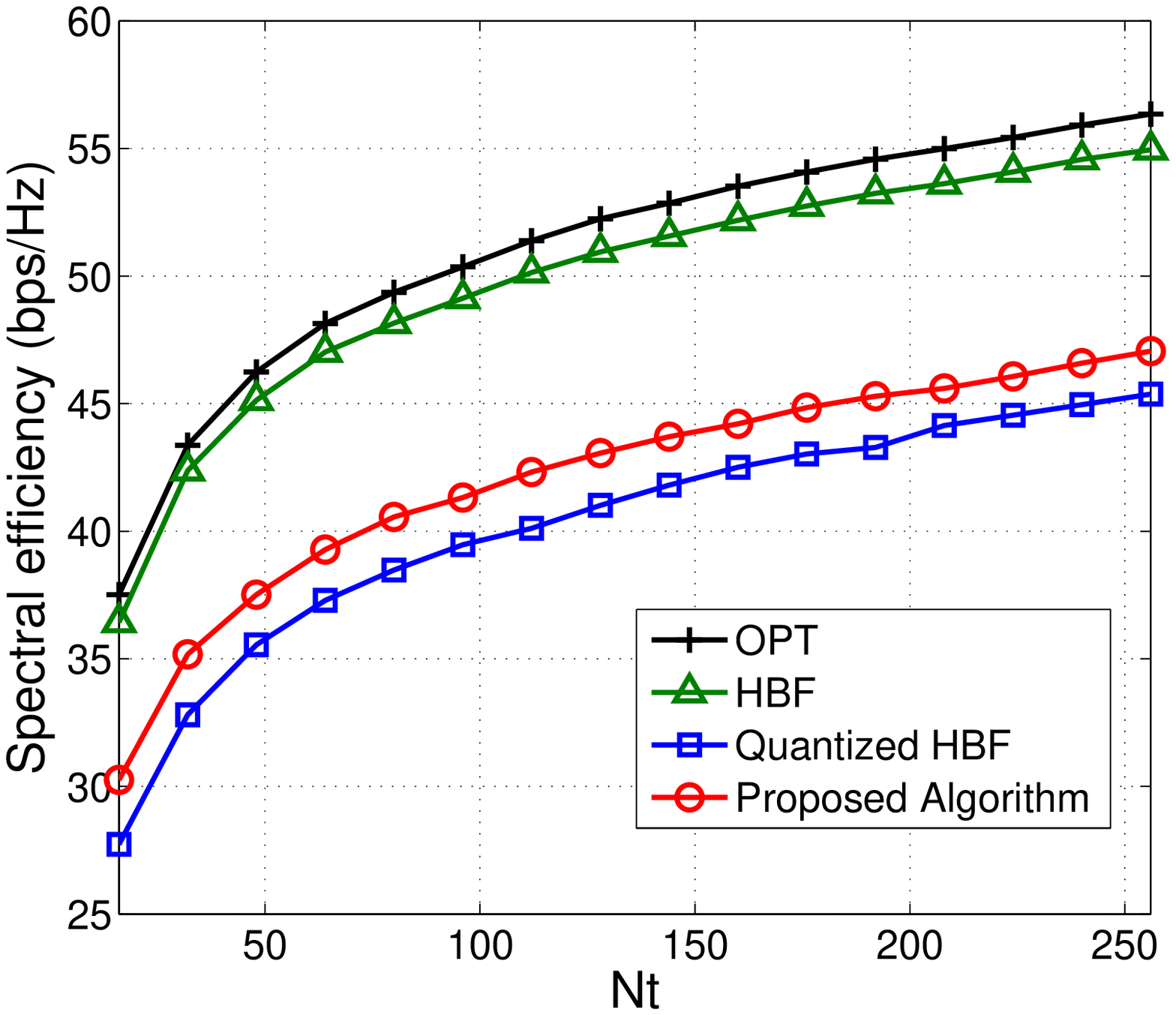}
  \vspace{-0.3 cm}
  \caption{Spectral efficiency versus $N_t$ ($N_r=16$, $N_t^{RF}= N_r^{RF}=4$, $N_s=4$, SNR = $20$dB).}\label{fig:rate_vs_Nt}\vspace{-0.0 cm}

\includegraphics[width=3.1 in]{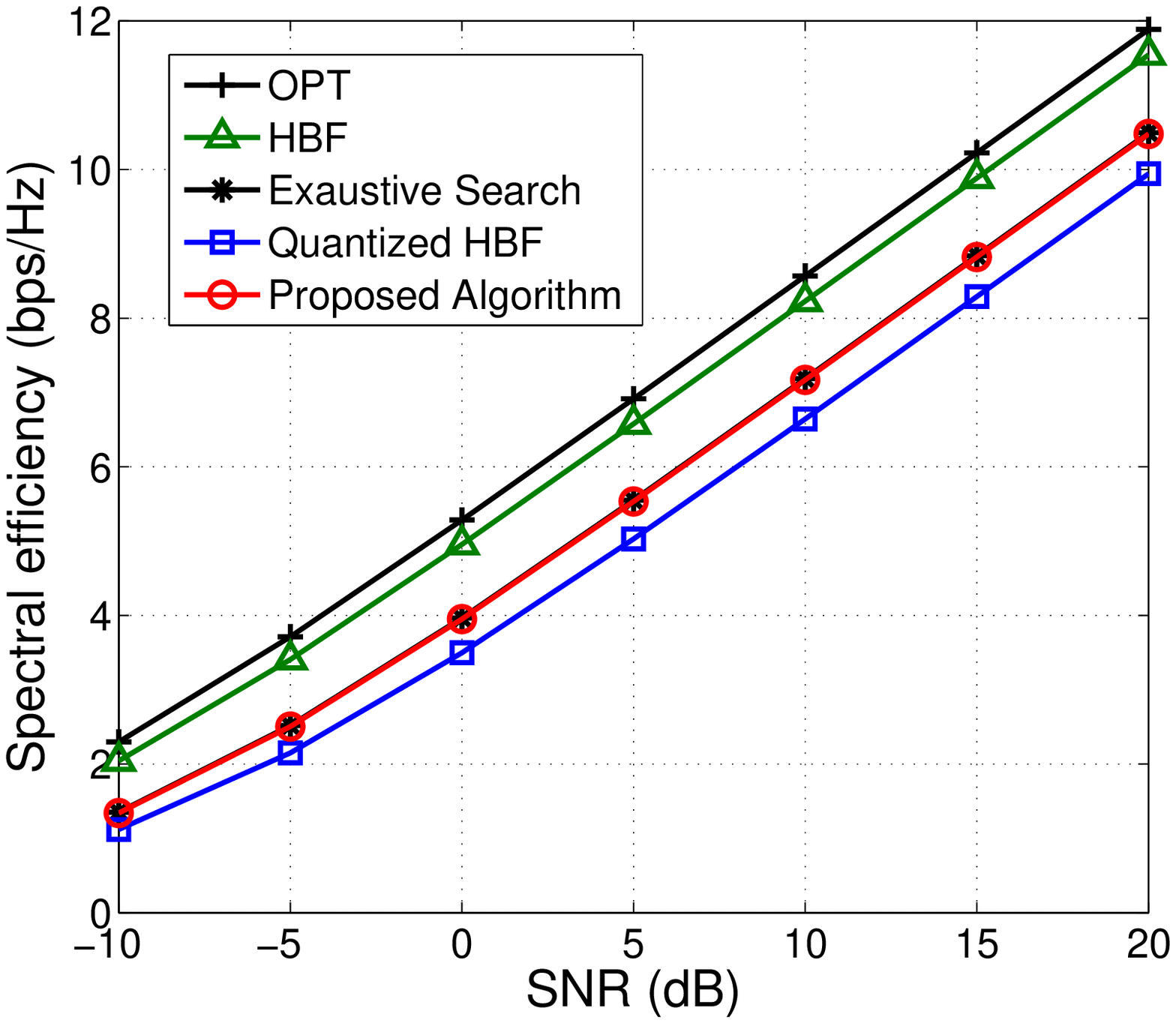}
  \vspace{-0.3 cm}
  \caption{Spectral efficiency versus SNR ($N_t= N_r=8$, $N_t^{RF}= N_r^{RF}=1$, $N_s=1$).}\label{fig:es_M8N8Ns1}\vspace{-0.4 cm}
\end{figure}

Fig. \ref{fig:rate_vs_Ns} provides spectral efficiency versus the number of data streams $N_s$. The number of transmit and receive RF chains are also changing along with $N_s$. We can see that our proposed algorithm can always outperform the Quantized HBF approach and the gap is increasing with the larger number of data streams.
In Fig. \ref{fig:rate_vs_Nt}, we turn to illustrate how the number of transmit antennas affects the spectral efficiency performance. The number of transmit antennas $N_t$ is varying from $16$ to $256$, while $N_r$ is fixed at $16$. The SNR is set at $20$dB and $N_t^{RF}=N_r^{RF}=N_s=4$. Similar conclusions can be drawn that the proposed algorithm has a notable performance advantage over the Quantized HBF approach.

Finally, in Fig. \ref{fig:es_M8N8Ns1} we investigate the performance loss between the one-bit quantized PSs based hybrid beamforming algorithms and the optimal exhaustive search method, where $N_t= N_r=8$ and $N_t^{RF}= N_r^{RF}=N_s=1$. It can be observed that the proposed algorithm can achieve almost the same performance compared to the exhaustive search scheme, verifying that our proposed algorithm can achieve near optimal solution with much lower complexity.

\section{Conclusions}
\label{sc:Conclusions}

This paper considered the problem of hybrid precoder and combiner design with one-bit quantized PSs in mmWave MIMO systems.
We proposed to firstly design the binary analog precoder and combiner pair for each data stream successively. We presented a novel binary analog precoder and combiner optimization algorithm under a Rank-1 approximation of the interference-included equivalent channel with lower than quadratic complexity.
Then, the digital precoder and combiner were computed based on the obtained baseband effective channel to further maximize the spectral efficiency.
Simulation results demonstrated the performance improvement of our proposed algorithm compared to the existing one-bit PSs based hybrid beamforming scheme.

\end{document}